\documentclass[aps,reprint,superscriptaddress]{revtex4-1}
\usepackage[dvipdfmx]{graphicx}
\usepackage{bm}
\usepackage{amsmath,amssymb}
\usepackage{physics}

\usepackage{epsf}
\usepackage[dvipdfmx]{graphicx}
\usepackage{color}
\usepackage{newfloat}

\begin{document}

\title{Appearance of ferromagnetism in Pt(100) ultrathin films originated from quantum-well states}

\author{Tatsuru Yamada}
\affiliation{Department of Applied Physics and Physico-Informatics, Keio University, Hiyoshi, Yokohama 223-0061, Japan}

\author{Keisuke Ochiai}
\affiliation{Department of Applied Physics and Physico-Informatics, Keio University, Hiyoshi, Yokohama 223-0061, Japan}

\author{Hirofumi Kinoshita}
\affiliation{Department of Applied Physics and Physico-Informatics, Keio University, Hiyoshi, Yokohama 223-0061, Japan}

\author{Shunsuke Sakuragi}
\email[e-mail: ]{shunsuke\_sakuragi@mri-ra.co.jp}
\affiliation{MRI Research Associates, Inc., 2-11-1 Nagata-cho Chiyoda, 100-6105, Japan}
\affiliation{ISSP, The University of Tokyo, Kashiwa, Chiba, 277-8581 Japan}

\author{Motohiro Suzuki}
\affiliation{Japan Synchrotron Radiation Research Institute, Kouto, Sayo 679-5198, Japan}

\author{Hitoshi Osawa}
\affiliation{Japan Synchrotron Radiation Research Institute, Kouto, Sayo 679-5198, Japan}

\author{Hiroyuki Kageshima}
\affiliation{Graduate School of Natural Science and Technology, Shimane University, Nishikawatsu-cho, Matsue 690-8504, Japan}

\author{Tetsuya Sato}
\email[e-mail: ]{satoh@appi.keio.ac.jp}
\affiliation{Department of Applied Physics and Physico-Informatics, Keio University, Hiyoshi, Yokohama 223-0061, Japan}

\date{\today}

\begin{abstract}
Ferromagnetism was observed in a Pt(100) ultrathin film deposited on a SrTiO$_3$(100) substrate. The ferromagnetism, which appears in films with thicknesses of 2.2-4.4 nm, periodically changes with a period of approximately 1 nm (5-6 monolayers) depending on the film thickness. This is consistent with the period derived from the quantum-well states formed in the thin film. X-ray magnetic circular dichroism measurements show the evidence of appearance of intrinsic ferromagnetism in Pt(100) ultra-thin film. In addition, present results suggest a possibility that the orbital magnetic moment of pure Pt is much smaller than that of the Pt/ferromagnetic multilayer system. 
\end{abstract}


\maketitle

\section{Introduction}
Pt and Pd are both nonmagnetic but, based on Stoner's standpoint, are considered to be nearly ferromagnetic in the bulk form. In other words, these metals have a large density of states at the Fermi energy $D(\epsilon_{\rm F})$ \cite{stoner}. 
In recent years, Pt and Pd have been reported to exhibit ferromagnetic properties at the nanoscale \cite{ShinoharaPRL, SampedroPRL, TaniyamaEPL, mirbtPRB, niklassonPRB, hongPRB, ObaXMCD,SakamotoPRB, SakuragiPRB}. 
This is because the change in the electronic structure, which occurs with the discreteness of the electronic state at the nanoscale, can significantly modulate $D(\epsilon_{\rm F})$. 
In particular, in ultrathin films, Pd(100) exhibits ferromagnetic behavior in an oscillatory manner. This behavior, which depends on the film thickness, is exhibited with a periodic increase in $D(\epsilon_{\rm F})$ owing to the formation of quantum-well (QW) states \cite{mirbtPRB, niklassonPRB, hongPRB, SakuragiPRB}. 
In addition, the magnetism in Pd(100) ultrathin films can be controlled by applying an external field to change the QW states \cite{sunPRB, aiharaJAP, BauerPRB, TanabeAPL}. 

Pt, an element that is homologous with Pd, is an important metal for spin devices because of its large spin-orbit coupling \cite{spinpomp, GuoPRL}. 
In particular, the magnetic anisotropy of alloys that contain Pt, such as $L1_{0}$ FePt, is known to be extraordinarily large \cite{FePt}. Many studies have attempted to explain the microscopic mechanism whereby the magnetism in Pt can be controlled by using external fields and/or the proximity effects of ferromagnetic materials\cite{ObaPRL, YamadaPRL, MiwaNcom, ShimizuPRL, Liangeaar2030}. Despite the interesting physics of this metal, few reports regarding the ferromagnetism of pure Pt are available.

In this paper, we report the discovery that Pt, which is a nonmagnetic metal in bulk form, exhibits ferromagnetism with large magnetization in (100) thin films on a SrTiO$_3$ (STO) substrate. The oscillation of the magnetization depends on the film thickness within the thickness range of 2.2-4.2 nm. In addition, the oscillation period is in good agreement with the first-principles calculation relating to the two-dimensional QW states in the electronic structure. Based on previous studies of QW-induced ferromagnetism, this suggests that the magnetism and band structure of the 5$d$ electrons in ultrathin films of Pt(100) can be controlled by external fields. X-ray magnetic circular dichroism (XMCD) spectroscopy of 3.17 nm thin film indicated that there is intrinsic ferromagnetism in Pt(100) ultrathin film. 

\section{Experimental details}
We prepared epitaxial Pt(100) ultrathin films using a molecular beam epitaxy chamber with a base pressure of $\sim 1\times 10^{-9}$ Torr on an STO substrate, which has a Ti-O-terminated step-and-terrace surface structure (SHINKOSHA Co., Ltd.) \cite{KawasakiScience}. 
Film growth consisted of the following three steps \cite{wagnerJAP}. First, one-fifth of the total thickness of the Pt film was deposited at 653 K. Second, the remainder was deposited at room temperature; finally, the deposited film was heated to $\sim$573 K, and this process was recorded by capturing reflection high-energy electron diffraction (RHEED) images. The RHEED image of the deposited Pt layer is shown in Fig. 1, in which the sharp streaks indicate the high crystallinity and atomic flatness of the Pt films \cite{PhysProc, SakuragiPRB2}. 
We note that, RHEED images like Fig. 1 cannot be obtained below thickness with 2.2 nm (the image which indicates island growth was observed). 

\begin{figure}
\centering
\includegraphics[width=6.5cm]{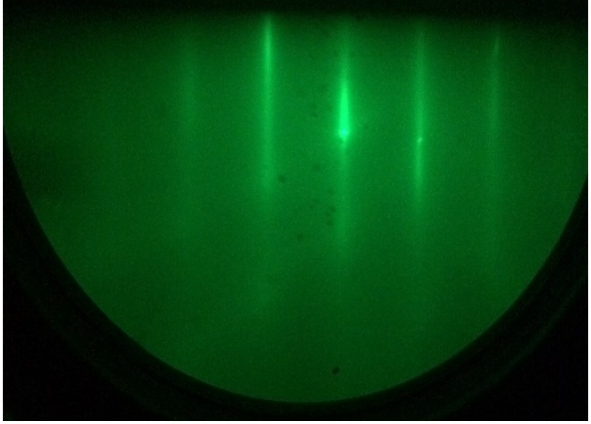}
\caption{\label{fig1} 
RHEED image of the Pt(100) ultrathin film on the STO(100) substrate. The streak lines indicate high crystallinity and atomic flatness.  
}
\end{figure}

The prepared samples were encapsulated in quartz tubes attached to the bottom of an ultra-high vacuum chamber to prevent gas adsorption \cite{SakuragiPRB}. Encapsulation of the samples in quartz tubes makes it possible to measure the magnetization with a superconducting quantum interference device (SQUID) magnetometer in the vacuum state. We note that the signal of ferromagnetic impurities are not observed from STO substrate and quartz tubes at least in the range of measurement sensitivity of SQUID. 

To measure the intrinsic magnetism of the Pt(100) ultrathin films, XAS and XMCD measurements were conducted on beamline BL39XU of the SPring-8 synchrotron radiation facility after the SQUID magnetometer measurement. 
Here, we used the photon-in photon-out fluorescence yield method with a circularly polarized hard X-ray beam around the Pt $L_3$ ($\sim$11.57 keV) and $L_2$ ($\sim$13.28 keV) edges. The incident angle of the X-rays was fixed at 10$^\circ$ and measurements were carried out in an atmosphere of non-active gas flow. 
Each measurement was conducted by employing a total of four types of polarization reversals using diamond X-ray phase plates and by reversing the magnetic field of the electromagnet to remove artifacts. 
The attenuation length of Pt during hard X-rays (~11 keV) is more than 1 micrometer, and present ~3 nm film is about 1/1000 compared to that length. Thus, the effect of self-absorption is negligible. 

All magnetometry measurements were performed at room temperature $\sim$ 300 K to avoid the effect of structural phase transition of the STO substrate, which induced the degeneration of the films \cite{SakuragiJMMM}. After the magnetometry experiments, the film thickness was evaluated by X-ray reflection measurement using a Cu $K\alpha$ laboratory source. 

\section{EXPERIMENTAL RESULTS}
Figure 2(a) shows the nonlinear components of the magnetization of Pt(100) ultrathin films, which were obtained by subtracting the diamagnetic moment of the STO substrate, measured at 300 K using a SQUID magnetometer. 
The remanent magnetization and coercive field of 100-200 Oe in Pt(100) ultrathin films, which have nonlinear magnetization curves, are plotted in Fig. 2(b).

\begin{figure}
\centering
\includegraphics[width=8cm]{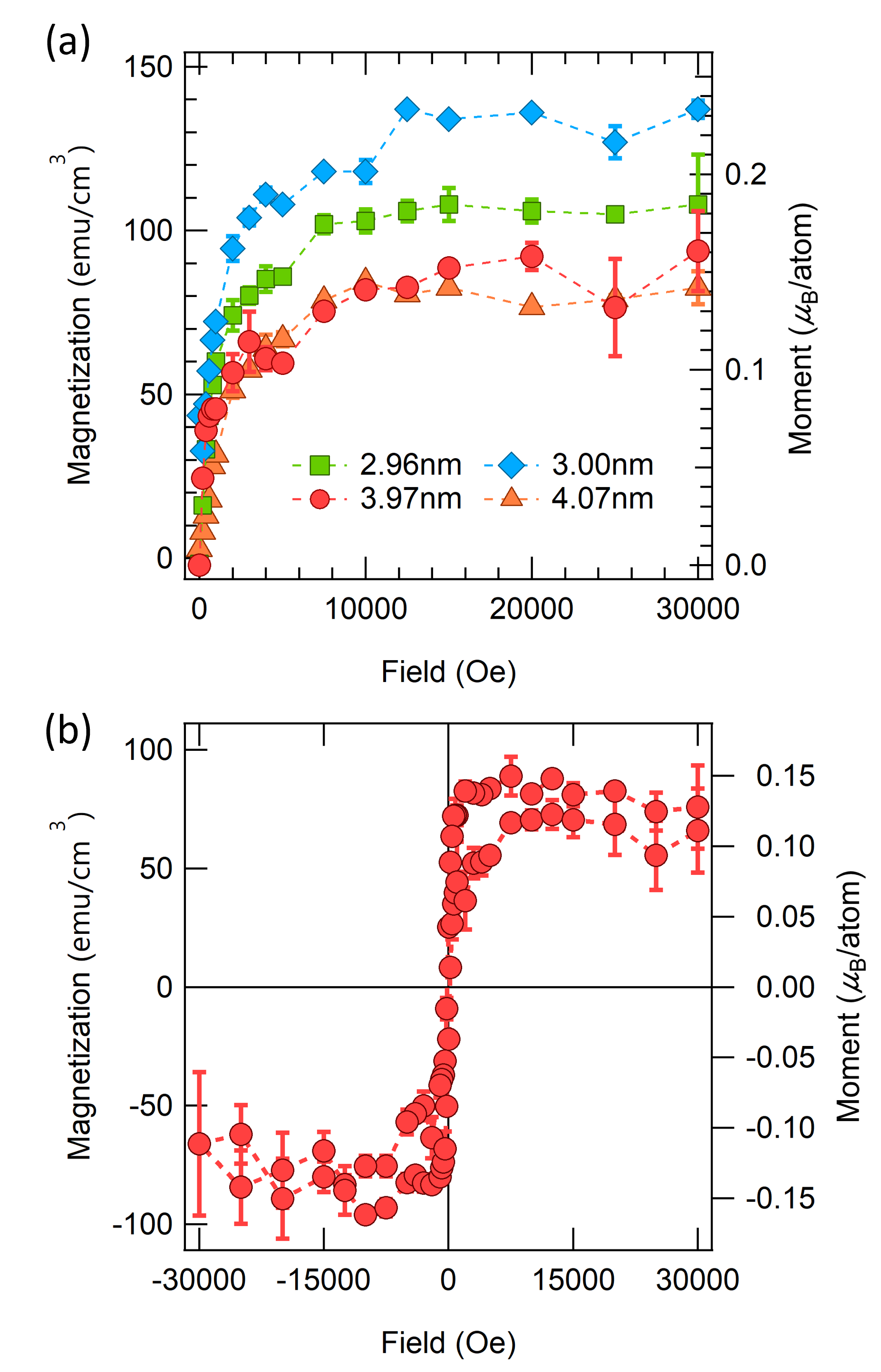}
\caption{\label{fig2} 
(a) Extracted nonlinear components of the magnetization of Pt(100) films measured at 300 K. (b) Nonlinear magnetization curves of a 3.97 nm thick film. The inset shows the coercive field of the film of $\sim$200 Oe. 
}
\end{figure}

To examine the magnetism of the Pt(100) ultrathin film in detail, X-ray absorption spectroscopy (XAS) and XMCD measurements were performed on a 3.17 nm thin film, which was exposed to air for a few minutes before the measurements. 
The XAS profiles recorded at an external magnetic field of 20,000 Oe corresponds with the position of the XMCD peak at this field at both the $L_3$ and $L_2$ edges [Fig. 3(a)-(c)]. 
In addition, the two peaks that were detected using XMCD and that appeared at the $L_3$ and $L_2$ edges have opposite signs. 
This is consistent with the XAS and XMCD data obtained for Pt, in which a ferromagnetic moment is induced owing to the proximity effect of ferromagnetic materials. 
Present obtained XMCD signal is over 10 times greater than that of Pt foil which shows Pauli paramagnetism \cite{Pauli,comment220818}.

\begin{figure*}
\includegraphics[width=17.2cm]{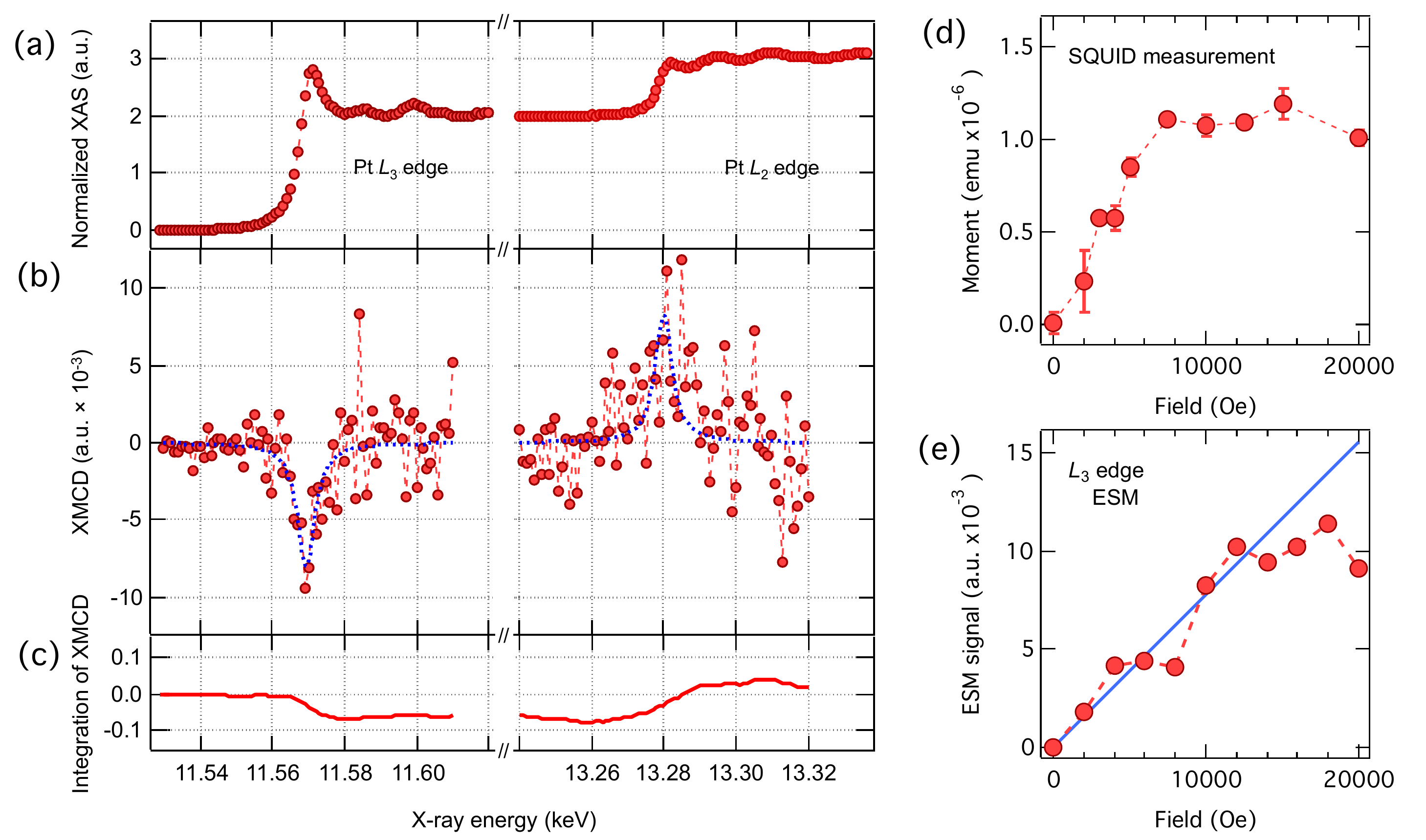}
\caption{\label{fig3}
(a) and (b) XAS and XMCD signals of the Pt(100) ultrathin film with a thickness of 3.17 nm taken at the Pt $L_3$ and $L_2$ edges. An external magnetic field of 20,000 Oe was applied. Blue lines indicate the Lorentz function (see Appendix A). 
(c) Integration of XMCD signals of the Pt(100) ultrathin film obtained form measurement plots (not using the Lorentz function). 
(d) Initial magnetization curve of Pt(100) ultrathin film obtained from the SQUID magnetometer. This magnetization curve was measured under vacuum before the synchrotron measurement, and the obtained magnetic moment was $\sim$0.06 $\mu_B$/atom. 
(e) ESM curve of the Pt(100) ultrathin film recorded at the Pt $L_3$ edge. The ESM values are inverted to allow them to be compared with the SQUID measurement. The straight line is a linear approximation of the ESM between 0 and 6000 Oe.
}
\end{figure*}

Figure 3 (e) shows the element-specific magnetometry (ESM) measurements using the Pt $L_3$ edge. 
The magnetization curve that was obtained by plotting these measurements saturates at approximately 12,000 Oe. 
This is qualitatively consistent with the initial magnetization curve of the same Pt(100) ultrathin film that was obtained using a SQUID magnetometer in Fig.3 (d) (the saturated field is $\sim 8,000$ Oe). 
We note that the direction of the applied magnetic field in SQUID magnetometer leans due to the sample size in Fig.3 (d), and direction of magnetic field is different between Fig.3 (d) and (e).  

To compare the XMCD and SQUID magnetometer measurements, we estimated the orbital magnetic moment $m_{orb}$ and the effective spin magnetic moment $m_{spin}^{eff}$ (=$m_{spin} - 7 \expval{T_{z}}$, where $\expval{T_{z}}$ is the expected value of the magnetic dipole operator) from the XMCD measurement using the XMCD sum rules \cite{TholePRL, CarraPRL}. 
As a result, each moment per Pt atom in the entire volume of the Pt(100) ultrathin film $m_{orb}$=0.0003 $\pm$ 0.0009 $\mu_B$ and 
$m_{spin}^{eff} = 0.0184 \pm 0.0022 \mu_B$ are obtained from the fit curves of Lorentz function; thus, the total magnetic moment $m_{tot}$ is 0.0186 $\pm$ 0.0023 $\mu_B$ (see Appendix A). 
From the SQUID magnetometer measurement, the magnetic moment of the sample was found to be $\sim$0.06 $\mu_B$, and the XMCD result is smaller than the SQUID result. 
This could be attributed to the deterioration of the surface of the Pt(100) ultrathin film as a result of the ozone generated by hard X-ray irradiation. 
In the case of Pt, Pd nano particles and Pd ultrathin films, ferromagnetic state is sensitive to the surface state and decrease of magnetic moment occurs by air and/or chemical adsorption \cite{ObaJPSJ, SakuragiPRB}. Thus, the present observed decrease in magnetic moment by ozone is consistent with the peculiarity of ferromagnetic Pt and Pd. However present XMCD contains large noises and the uncertainty still remains of present obtained magnetic moment from XMCD, the large amount of XMCD signal indicates that there is intrinsic ferromagnetism in Pt(100) ultrathin film. 

Based on the XMCD sum rules, the ratio of the orbital magnetic moment to the effective magnetic moment $m_{orb}$/($m_{spin}^{eff}$) is estimated to be $0.015 \pm 0.046$. 
This value is much smaller than the ratios obtained for Pt in which ferromagnetism is induced by the proximity effect with ferromagnetic materials (0.14 for Co/Pt \cite{SuzukiPRB} and 0.087 for YIG/Pt \cite{KikkawaPRB}) and paramagnetism of bulk Pt (0.38 for Pt foil \cite{Pauli}).
This provides a potential to a possibility that the spin-orbit coupling in pure ferromagnetic Pt cannot generate a large amount of orbital momentum. 
Detailed experiments would be necessary to confirm the above because our present analysis is carried out for XMCD signal with a lot of noises. 

\section{DISCUSSION}
The discussion of the mechanism underlying the appearance of ferromagnetism in the present system, which consists of epitaxial Pt(100) ultrathin films, is based on Fig. 4(a), in which the magnetic moment per Pt atom, obtained from the spontaneous magnetization at 300 K in vacuum, is plotted as a function of the film thickness. 
The oscillatory change in the magnetic moment, with a period of $\sim$1 nm corresponding to 5-6 monolayers (ML), is clearly demonstrated in the range of 0-0.2 $\mu_B$/atom.
The error bar of magnetic moment originates from non-uniformity in film thickness within a sample \cite{SakuragiPRB2}.

\begin{figure}
\centering
\includegraphics[width=8cm]{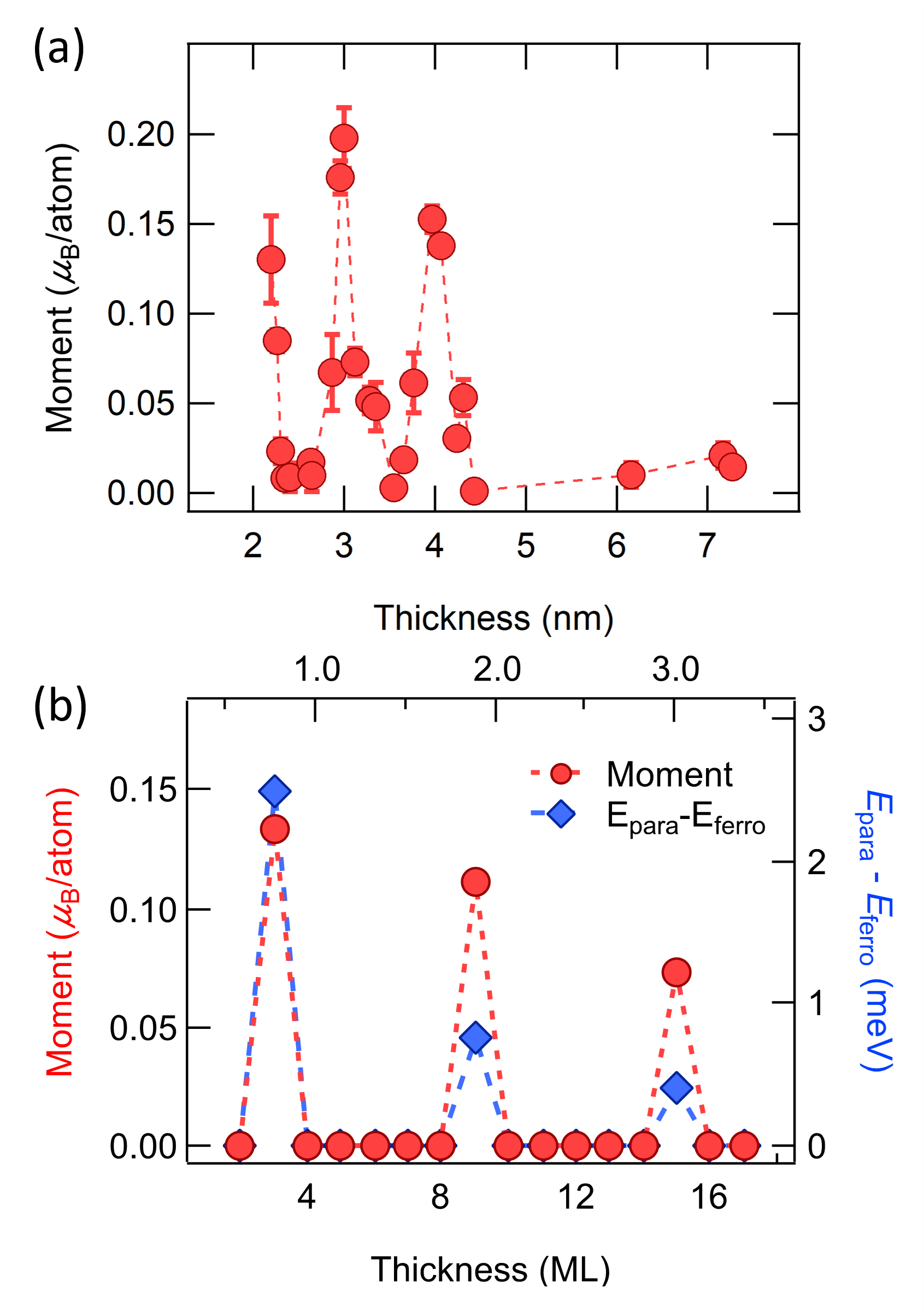}
\caption{\label{fig4} 
(a) Thickness-dependent magnetic moment per Pt atom, obtained from the saturation magnetization at 300 K. The oscillatory period of the magnetic moment is almost 1 nm, corresponding to thick films(5-6 ML). The raw data is available in Appendix B. The thickness is obtained by X-ray reflectivity measurement, thus we cannot obtain the number of atoms directly. (b) First-principle calculation of magnetic moment per Pt atom in Pt(100) thin films (circles) and total energy differences between the paramagnetic (para) and ferromagnetic (ferro) states (diamonds). The experimental period of oscillation in the magnetic moment is consistent with the calculated value. 
}
\end{figure}

To examine the magnetism in Pt(100) thin films in detail, we calculated the magnetic states of freestanding Pt(100) slabs by density functional theory (DFT) calculations using the generalized gradient approximation \cite{GGA}- projector augmented wave (PAW) method \cite{PAW} with $88 \times 88 \times 1$ $k$-points and a cutoff energy of 36 Ry (see Appendix C). 
We note that the present calculations were performed by ignoring the spin-orbit interaction because the orbital magnetic moment, obtained from the XMCD measurement, is small. 
For the above conditions, we evaluated the magnetic states of freestanding Pt(100) from 2 to 17 ML (from $\sim$0.4 to $\sim$3.4 nm), as shown in Fig. 4(b). 
Based on the DFT calculation, it was observed that the ferromagnetism oscillates depending on the film thickness, and the period of oscillation is consistent with our experimental observations. 
In particular, the ferromagnetism appears at a thickness of $\sim$3 nm (15 ML). This is in good agreement with the experimental result shown in Fig. 4(a).

Magnetism of an oscillatory nature that depends on the film thickness, such as the magnetism observed in this study, has also been reported for (100)-oriented thin films of Pd, which is also a 4$d$ transition metal \cite{SakuragiPRB}. 
Thus, it is suggested that the phenomenon of oscillation is a universal property of transition metals with $d$-electrons.  
In the case of Pd, the dependence of the oscillatory behaviour on the film thickness was not only discussed from the viewpoint of magnetometry, but also by considering the crystal growth and theoretical analysis of the band dispersion of $d$-electrons \cite{SakuragiPRB2, SakuragiPRB3}. 
Consequently, these phenomena have been well explained by the change in $D(\epsilon_{\rm F})$ owing to the formation of the $d$-electron QW states. 
We note that the band dispersion originated from $d$-electron QW states also observed at present DFT calculation. 

The surface and interface of the prepared ultrathin film behave as an energy barrier, which restricts the movement of electrons in the perpendicular direction and forms QW states. The electronic state can be described by the wave number $k_z$ based on the phase model \cite{ortegaPRB, chiang, SakuragiPRB3}, 
\begin{equation}\label{Phase}
2 \pi k_z N + (\phi_s(\epsilon, k_{x,y}) + \phi_i(\epsilon, k_{x,y}) ) = 2 \pi j,
\end{equation}
where $N$ is the number of layers, $\phi_{s, i}$ are the correction terms for the shift in the scattering phase at the surface and interface, and $j$ is an integer quantum number. 
If $\Phi(\epsilon, k_{x,y}) = -(\phi_s + \phi_i )/2\pi$, then $k_z=( j+\Phi)/N$ in units of $2\pi/a$, where $a$ is the lattice constant. 
The period $p$ at which the discrete band overlaps with the Fermi energy is evaluated as follows using the Fermi wave number $k_F$ when $k_F >0.5$: 
\begin{equation}\label{period}
p = 1 /(1 - k_F ).
\end{equation}
Generally, the band structures of QW states represent the bulk electronic states \cite{DabrowskiPRL}. 
Thus, in Eq. (\ref{period}), we use the Fermi wavenumber $k_F$ of the bulk Pt(100) values, which originates from the 5$d$ electrons, and $p$ is estimated to be 5.7 ML \cite{AndersonPRB}. 
This is consistent with our experimental and calculated results. 
Therefore, we determined that $D(\epsilon_{\rm F})$ increases when the discretized band according to the QW states overlaps with the Fermi energy, whereupon the Pt(100) thin film becomes ferromagnetic in terms of the Stoner criterion.

The results of our DFT calculations shown in Fig. 4(b) underestimate the value of the magnetic moment. 
The same discordance between the experimental and DFT results was reported for QW-induced ferromagnetism in Pd. 
In the case of Pd, the QW-induced lattice expansion enhances the stability of the ferromagnetic state \cite{SakuragiPRB2, PhysProc}. 
This lattice distortion in the films could explain the discrepancy in the amplitude of the magnetic moment of a Pt atom between the experimental and calculated data \cite{KanaPRB, EkmanPRB, BanAPL}. %

Here, we note that the scattering phase shifts $\Phi$ in Eq. (\ref{Phase}) depend on the in-plane wave number $k_{x,y}$ \cite{SakuragiPRB3}. 
This indicates that the shape of the discretized band dispersion, i.e., magnetization depends on the shape of the function $\Phi$, that is, electronic interaction at the Pt/substrate interface. 
A comparison between the experimental and DFT results indicates that the value of the phase shift in Pt(100)/STO is smaller than that of Pd(100)/STO, which was reported previously \cite{SakuragiPRB}. This indicates that the electronic interactions at the film/substrate interfaces of Pt and Pd are different. 
The lattice mismatch between Pd/STO (-0.04 \%) and Pt/STO (0.46 \%) are different, and thus, the layer distance at STO interface should be different between Pd and Pt. This can change the electronic interaction at the STO interface. 
A detailed discussion based on the energy band calculation, surface X-ray diffraction, and angle-resolved photoemission spectroscopy experiments would be necessary to elucidate the effect of the STO substrate on the appearance of ferromagnetism in Pt(100) ultrathin films. 
Nevertheless, the QW mechanism is appropriate for explaining the dependence of the oscillatory magnetization on the film thickness in the Pt(100)/STO system.

\section{Conclusion}
Our experiments and calculations systematically demonstrated the appearance of ferromagnetism in Pt, which is a nonmagnetic metal in the bulk form. Pt becomes ferromagnetic according to the Stoner criterion as a result of the QW states formed in nanosized structures. 
Our XMCD experiment showed that the appearance of intrinsic ferromagnetism in Pt(100) films. 
The Pt(100) ultrathin film system is expected to be useful to adjust the state of the Pt 5$d$ electrons by changing the film thickness and/or by applying an external field to modify the QW states. 
This suggests the concept for controlling of the magnetism of multilayer Pt(100)/ferromagnetic materials, which have large magnetic anisotropy. 
Our present findings provide an innovative platform for discussions on ways to artificially create and freely control magnetic materials.

\section*{ACKNOWLEDGMENTS}
We thank K. Mochihara and S. Yamaguchi for helpful suggestions on the theoretical calculations. The synchrotron radiation experiments were performed at the BL39XU beamline of SPring-8 with the approval of the Japan Synchrotron Radiation Research Institute (JASRI) (Proposal No.2019B1258). The computation in this work was performed using the facilities of the Supercomputer Center at the Institute for Solid State Physics of the University of Tokyo. This work was supported by the Japan Society for the Promotion of Science (JSPS) Grants-in-Aid for Scientific Research Grant No. 19K051999

\appendix
\section{XMCD sum rules analysis}
From the XMCD measurement, we estimated the orbital moment ($m_{orb}$) and effective spin moment ($m_{spin}^{eff}$) of the Pt(100) ultrathin film based on the following sum rules \cite{TholePRL, CarraPRL}: 
\begin{equation}\label{Sum1}
m_{orb} = -\frac{2}{3} \frac{\Delta I_{L_3} + \Delta I_{L_2} }{I_{L_3} + I_{L_2}} n_h \mu_B, 
\end{equation}
\begin{equation}\label{Sum2}
m_{spin}^{eff} = - \frac{\Delta I_{L_3} - 2 \Delta I_{L_2}}{I_{L_3} + I_{L_2}} n_h \mu_B, 
\end{equation}
where $I_{L_3}+I_{L_2}$ is the integrated summed value of the white line signals of XAS over the Pt $L_{3, 2}$ edges, $\Delta I_{L_{3,2}}$ is the integral of the XMCD spectra for the Pt $L_{3, 2}$ edges, $n_h$ is the number of holes in the Pt 5$d$ bands, and $\mu_B$ is the Bohr magneton. 
To subtract the background signal and contribution from electron transitions to the continuum in the raw XAS spectra, we used the inverse tangent step function with linear components. After subtraction, the XAS patterns were fitted using the Lorentz function, and the white line spectra of XAS $L_{3, 2}$ were obtained. 
The integrated values of the XMCD profiles $\Delta I_{L_{3, 2}}$ were obtained by fitting with the same Lorentz function as in the XAS analysis; using same half value of width and peak position. 
Previously, $n_h = 1.89$ was reported for the 5$d$ bands of a Pt film\cite{KikkawaPRB}, and we used this value for the present analysis. 
Based on the above, we obtained $m_{orb}$=0.0003 $\pm$ 0.0009 $\mu_B$, $m_{spin}^{eff} = 0.0184 \pm 0.0022 \mu_B$, and the total magnetic moment $m_{tot}$ is 0.0186 $\pm$ 0.0023 $\mu_B$ in the present XAS and XMCD measurements of the Pt(100) ultrathin film in Figs. 3.

\section{Thickness-dependent magnetic moment of Pt(100)}
Table \ref{tab1} shows the raw data of thickness-dependent magnetic moment of Pt(100) ultrathin films, shown in Fig. \ref{fig4}(a). 

\begin{table}[]
\caption{\label{tab1} Thickness-dependent magnetic moment of Pt(100)}
\begin{tabular}{ccc}
\hline
Thickness (nm) & Moment ($\mu_B$/atom) & Error ($\mu_B$/atom) \\
\hline \hline
2.20           & 0.13                           & 0.02                           \\
2.27           & 0.09                           & 0.01                           \\
2.30           & 0.02                           & 0.01                           \\
2.35           & 0.01                           & 0.00                           \\
2.41           & 0.01                           & 0.01                           \\
2.63           & 0.02                           & 0.00                           \\
2.65           & 0.01                           & 0.01                           \\
2.87           & 0.07                           & 0.02                           \\
2.96           & 0.18                           & 0.01                           \\
3.00           & 0.20                           & 0.02                           \\
3.12           & 0.07                           & 0.01                           \\
3.28           & 0.05                           & 0.01                           \\
3.35           & 0.05                           & 0.01                           \\
3.55           & 0.00                           & 0.00                           \\
3.66           & 0.02                           & 0.00                           \\
3.77           & 0.06                           & 0.02                           \\
3.97           & 0.15                           & 0.01                           \\
4.07           & 0.14                           & 0.00                           \\
4.24           & 0.03                           & 0.01                           \\
4.31           & 0.05                           & 0.01                           \\
4.44           & 0.00                           & 0.00                           \\
6.16           & 0.01                           & 0.01                           \\
7.17           & 0.02                           & 0.01                           \\
7.28           & 0.01                           & 0.00                           \\
\hline 
\end{tabular}
\end{table}

\section{Calculation details}
The PHASE/0 program \cite{PHASE} using the projector augmented wave (PAW)\cite{PAW} was used to calculate the magnetism in Pt(100) ultrathin films. The generalized gradient approximation (GGA) reported by Perdew, Burke, and Ernzerhof \cite{GGA} was used for the density functional theory calculations. The value of the lattice constant converges to 0.397 nm for fcc bulk Pt, and this value was adopted for the film-shaped Pt(100). To evaluate the magnetism of Pt(100) ultrathin films, a slab of vacuum(two monolayers)/Pt(N monolayers)/ vacuum (three monolayers), $88 \times 88 \times 1 k$-points, and cutoff energy of 36 Ry were used. The dimensions of the lateral unit cell were $1 \times 1$. Based on this, we calculated the difference in the total energies between the paramagnetic and ferromagnetic states, where the spin polarization was assigned a certain constant value in freestanding Pt(100). The calculations were performed by ignoring the spin-orbit interaction because the orbital magnetic moment, obtained from XMCD data, is very small.

\bibliography{ref}

\end{document}